\documentclass[allclo,superscriptaddress,eqsecnum,amsfonts,showpacs]{revtex4}
\usepackage{epsfig}

\newcommand{\be}{\begin{equation}}
\newcommand{\ee}{\end{equation}}
\newcommand{\bea}{\begin{eqnarray}}
\newcommand{\eea}{\end{eqnarray}}

\newcommand{\ep}{i\varepsilon}
\newcommand{\nn}{\nonumber}

\begin{document}

\preprint{ \parbox{1.5in}{\leftline{hep-th/??????}}}

\title{Confinement of electrons in QED2+1 and quarks in QCD3+1 in Temporal Euclidean space}

\author{Vladimir ~\v{S}auli}
\affiliation{CFTP and Dept. of Phys.,
IST, Av. Rovisco Pais, 1049-001 Lisbon,
Portugal }

\begin{abstract}
Without any analytical assumption we solve the ladder QED2+1 in Minkowski space.  
Further, we transform Greens functions  
to the Temporal Euclidean space, wherein we  show that in the special case of ladder 
QED2+1 the solution  is fully equivalent to the  Minkowski one. QCD quark gap equation is solved in the framework of Temporal Euclidean space as well.  
In both models, the obtained complex fermion propagators exhibits confinement,
since it does not satisfies Khallen-Lehmann representation.

\end{abstract}

\maketitle
%

\section{Introduction}

 Quark confinement in Quantum Chromodynamics (QCD) is a phenomenon of current
interest. Due to the fact that QCD is not easily tractable, the toy models which exhibit QCD low energy phenomena --the confinement and chiral symmetry breaking-- are often investigated.
In this respect 2+1d Quantum Electrodynamics (QED2+1) has these similarities with QCD in the
usual 3+1d Minkowski space.  
Based on the Euclidean space study of QED3, the chiral symmetry breaking for a small number flavors has been proposed  for the first time in \cite{APNAWI1988}. Since the scale of dynamical chiral symmetry breaking, being characterized  by a fermion mass in the infrared -$M(0)$-, is
one order of magnitude smaller then the topological dimensioned coupling $e^2$, the 
Schwinger-Dyson equations (SDEs) provide a unique powerful framework for the nonperturbative study, see e.g. most recent studies \cite{BARA2007,BARACORO2008}.  Particularly, confinement in relation with  dynamical  complex pole generation in fermion propagator has been discussed in 
\cite{MARIS1995}.

However, the all aforementioned  studies have been done in the
standard Euclidean
space. That is, after performing the standard Wick rotation \cite{WICK} of the timelike
components of the momentum variables (internal integral momentum as well as external one, explicitly $p_3^{E}=-ip_0^{M}$,  the measure $id^3p^{M}=-d^3p^{E}$). It is assumed and widely believed, that the Green`s functions for timelike arguments can be obtained after analytical continuation of the functions calculated in  Euclidean space, where it is supposed the Euclidean solution itself  represent the correct Minkowski solution for the spacelike arguments. Therefore, to shed a new light and for the first time,  we solve fermion QED2+1 SDE directly in 2+1 Minkowski space. The so called ladder approximation of electron SDE is introduced in the the Section II.

The equivalence of Minkowski solution with the so called Temporal Euclidean space has been proved in 
\cite{SAZA2009}. Recall here,  $E_T$  space  metric is obtained from Minkowski one by the multidimensional  Wick rotations, but now instead for the time component, it is  made for  all the space coordinates of the Lorentz three vector \cite{SAZO2008}. Lately, the method of multidimensional Wick rotation has been applied to QCD light quark sector and the solution ha been found for the u,d quarks gap equation.

In this talk , the numerical results for various ratio of the coupling and the electron mass 
are presented for QED2+1 and Temporal Euclidean QCD. 
  The resulting  propagators describe "propagation" of confined object, they have no real poles and they violates reflection positivity- they do not have Khallen-Lehmann representation-. The appropriate excitations  can never be on-shell and thus never observed as a free particles. In the case of QED2+1 the (in-)validity of standard Wick rotation was discussed in \cite{SAZA2009}.

The Schwinger-Dyson equations are presented for QED2+1. For QCD quark SDE we just mention the detail of the model and present the results.

\section{Direct Minkowski space solution of QED2+1 in ladder approximation}

In our study we employ Minkowski metric $g_{\mu\nu}=diag(1,-1,-1)$, in order to  properly describe chiral symmetry, we use the standard four dimensional Dirac matrices such that they anticomutation relation is  $\left\{\gamma_{\mu},\gamma_{\nu}\right\}=2g_{\mu\nu}$. With these conventions the inverse of the full fermion propagator reads
\bea \label{gap}
S^{-1}(p)&=&\not p- m-\Sigma(p)
\nn \\
\Sigma(p)&=&ie^2\int \frac{d^3k}{(2\pi)^3} \, , G^{\mu\nu}(k-p)\Gamma_{\mu}(k,p)S(k)\gamma_{\nu}
\eea

We consider the explicit chiral symmetry breaking  mass term of the form $m\bar{\psi}\psi$
so  parity is conserved. In this case the dressed fermion propagator can be parametrized
by two scalar function like
\be
S(p)=S_v(p)\not p+S_s(p)=\frac{1}{\not p A(p)-B(p)}\, . 
\ee

The full photon propagator $G$ and the electron-positron-photon vertex $\Gamma$ satisfy their own
SDEs.  

The ladder approximation is the simplest selfconsistent approximation which approximate  the unknown Greens functions be their free counterpartners, i.e. $\Gamma_{\mu}=\gamma_{\mu}$ and the photon propagator in linear covariant gauges is
\be
G_{\mu\nu}=\frac{-g_{\mu\nu}+(1-\xi)\frac{k_{\mu}k_{\nu}}{k^2}}{k^2} \, .
\ee

In general  QFT the Greens functions are not  real functions but complex tempered distributions.
In perturbation theory these are just real poles (together with its Feynman $\ep $ prescription) of the propagators, which when coincide in the loop integrals, produce branch cut starting at the usual production threshold. At one scalar loop level, the two propagators make the selfenergy complex above  the point $p^2=(M_1+M_2)^2$, wherein $M_1,M_2$ are the real masses- in fact the positions of these poles. Depending on the masses of the interacting fields, the real pole persists   when  situated  below the threshold or we get non-zero width and the free particle becomes resonance with finite lifetime.

 In strong coupling quantum field theory the mechanism of complexification can be very different (however the mixing of both mechanisms is not excluded). 
Here we simply assume that there is no zero at the inverse of propagator for real $p^2$, thus 
$\ep$ factor is not necessary and we integrate over the hyperbolic angles of Minkowski space directly.
For this purpose we have to consider the propagator function as the complex one for all $p^2$.
 A convenient parametrization of the  complex fermion propagator functions $S_s$ and $S_v$  can be written as
\bea
S_s(x)&=&\frac{B(k)}{A^2(k)k^2-B^2(k)}
\nn \\
&=&\frac{R_B\left[(R_A^2-\Gamma_A^2)k^2-R_B^2-\Gamma_B^2\right]+2R_A\Gamma_B\Gamma_A\,k^2}{D}
\nn \\
&+&i\, \frac{\Gamma_B\left[(R_A^2-\Gamma_A^2)k^2+R_B^2+\Gamma_B^2\right]-2R_BR_A\Gamma_A\,k^2}{D} \, ,
\label{ss}
 \\
S_v(k)&=&\frac{A(k)}{A^2(k)k^2-B^2(k)}
\nn \\
&=&\frac{R_A\left[(R_A^2+\Gamma_A^2)k^2-R_B^2+\Gamma_B^2\right]-2R_B\Gamma_A\Gamma_B}{D}
\nn \\
&+&i\, \frac{\Gamma_A\left[-(R_A^2+\Gamma_A^2)k^2
-R_B^2+\Gamma_B^2\right]+2R_A R_B\Gamma_B}{D}\, \, ,
\label{sv}
\eea

where $R_A,R_B$ $(\Gamma_A,\Gamma_B)$ are the real (imaginary) parts of the functions $A,B$
and the denominator $D$ reads
\be
D=([R_A^2-\Gamma_A^2] k^2-[R_B^2-\Gamma_B^2])^2+4(\Gamma_A R_A-\Gamma_B B)^2 \,.
\ee

In order to be able to compare between Minkowski and lately  considered Euclidean  spaces, the transformation in use should leave the Minkowski spacetime interval 
\be
s=t^2-x^2-y^2
\ee
manifestly apparent (i.e. $s$ could be a variable of the integral SDEs).   

To achieve this we will use 2+1 dimensional pseudospherical (hyperbolic) transformation of 
Cartesian Minkowski coordinates.
The obstacles followed by Minkowski hyperbolic angle integrals when going beyond $A=1$ approximation restrict us to the Landau gauge wherein the $A=1$ is the exact result in Temporal Euclidean space.

In momentum space our convenient choice of the substitution is the following:

\bea \label{myway}
\int d^3k K(k,p)=&&\int_0^{\infty} dr r^2 \int_0^{2\pi} d \theta  \int_0^{\infty} d \alpha
\left\{\sinh\alpha \, {\mbox{\begin{tabular}{|c|}
$k_o=-r\, \cosh\alpha $ \\ $k_x=-r \,\sinh\alpha \, \sin\theta$ \\
$k_y=-r \,\sinh\alpha\, \cos\theta$ \\
\end{tabular}}}
+\sinh\alpha\, {\mbox{\begin{tabular}{|c|} $k_o=r\, \cosh\alpha$\\ $k_x=r \,\sinh\alpha\, \sin\theta$\\ $k_y=r \, \sinh\alpha\, \cos\theta$
\end{tabular}}}\right. 
\nn \\
&&+\left.\cosh\alpha\, {\mbox{\begin{tabular}{|c|} $k_o=-r \sinh \alpha$\\$ k_x=-r \cosh\alpha\, \sin\theta$\\$ k_y=-r \cosh\alpha\, \cos\theta$
\end{tabular}}}+
 \cosh\alpha\, {\mbox{\begin{tabular}{|c|} $k_o=r \sinh \alpha$\\$ k_x=r \cosh\alpha\, \sin\theta$\\ $k_y=r \cosh\alpha\, \cos\theta$
\end{tabular}}}\right\} \, K(k,p) \, .
\eea

Notice, the integral boundaries are universal for  all the subregions of Minkowski space,
  the first line corresponds to the integration over the timelike 2+1momentum where we have 
\be
k^2=k_o^2-k_x^2-k_y^2=r^2>0 \, ,
\ee
where the left term corresponds to the negative energy interval $k_0<-\sqrt{k_x^2+k_y^2}$
and the right term corresponds to the positive  $k_0>+\sqrt{k_x^2+k_y^2}$. The second line
stands
 for the spacelike regime of the integration
\be  
k^2=-r^2<0\, ,
\ee
wherein the left term corresponds to the energy component interval  $k_0= (-\sqrt{k_x^2+k_y^2},0)$, while the right term in the second line stands for positive $k_0= (0,\sqrt{k_x^2+k_y^2})$ subspace of the full 2+1 dimensional Minkowski space. Functions $V$ in Rel. (\ref{myway}) represents the integrand of SDE.

The functions $A,B$  are Lorentz scalars, thus they can depend on $p^2$ only. We freely take the simple choice  of timelike external momenta as $p_{\mu}=(p,0,0)$. Integrating over the angles we can see that at the level of our approximation, the SDE separate for spacelike and timelike regime of the threemomenta. For timelike $p$ we get for the function $B$
\be \label{ET1}
B(p)=m+i(2+\xi)\frac{e^2}{4\pi^2}\int_0^{\infty} dk
\frac{k}{p}\ln\left|\frac{k+p}{k-p}\right|S_s(k) \, ,
\ee
where $\xi$ is a gauge parameter.
Stressed here, the Eq. (\ref{ET1}) is derived without any requirement of analyticity for the the propagator and the kernel.

For  the external spacelike Lorentz three-vector of momenta the $\alpha$ integration over the spacelike regime gives zero. For spacelike arguments the integration over the angle can not be performed analytically.

Let us  stress the main difference when compared to the standard  treatment. Here this is the timelike part of Minkowski subspace where the results are most naturally obtained.
Quite opposite to the standard approach where Minkowski solution is constructed by continuation of the Euclidean result, here the solution for the spacelike argument is non-trivially made from the timelike solution which must found as a first.

The validity of standard Wick rotation is highly speculative topics in the literature. Here we know the solution directly in Minkowski space and  the comparison with the Euclidean solution is straightforward and easy task. 
 Assuming the Wick rotation is valid, we could get the Minkowski solution equal to the Euclidean one at spacelike domain of momenta. We anticipate here and it will be explicitly shown in the next Section that the mass function $B$ is actually complex  for all real timelike $p^2$ at large window of studied parameters $m$ and $e$. Hence the Minkowski solution $B(0)$ being complex, is not the one literally known from the Euclidean studies, where $B_E(0)$ is an always purely real number.  These solutions do not coincide in the Minkowski light-cone (Euclidean zero) and in other words: commonly used strategy based on the analytical continuation of the Euclidean (spacelike) solution  to the timelike axis is wrong and should be abandoned in the case of QED2+1 theory.

In the paper \cite{SAZA2009} the proof of an equivalence of  QED fermion SDE formulated in Minkowski and Temporal Euclidean space has been shown. 
In the paper \cite{SAZO2008} it was proposed that N-dimensional analog of Wick rotations
performed for space components of Minkowski $N+1$-vector can be partially useful for  the study of a strong coupling quantum field theory.
In even dimensional  3+1QCD it is  the nonperturbative mechanism responsible for complex mass generation which is responsible for the absence of real pole type singularities in Greens functions evaluated at real their arguments which thus makes the nonperturbative calculations feasible there. 

In odd dimensional theory, like QED2+1 we study here, the complexification of masses and couplings can be quite naturally expected  because presence of  $i$  in the measures of the integrals defined in ET space. 
 Deforming the contour appropriately then the aforementioned-mentioned generalized Wick rotation  gives the following prescription for the momentum measure:
\bea
&&k_{x,y} \rightarrow i k_{2,3} \, ,
\nn \\
&&i \int d^3k \rightarrow  -i\int d^3k_{E_T} \, ,
\eea
 which, contrary to our standard  $3+1$ space-time, leaves the additional $i$ in front. 

In $E_T$ space  the singularity of the free propagator remains,   
for instance the free propagator of scalar particle is
\be
\frac{1}{p^{2}-m^2+\ep} \, ,
\ee 
with a positive square of the  three-momenta 
\be
p^2=p_1^2+p_2^2+p_3^2 \, ,
\ee
 thus formulation of the weak coupling (perturbation) theory, albeit possible, would not be more helpful then the standard approach (Wick rotation).

The advantage of the transformation to $E_T$ space becomes manifest, since the  fixed square Minkowski momentum $p^2=const$ hyperboloid with infinite surface is transformed into the finite  3dim-sphere in $E_T$ space. The Cartesian variables are related to the  spherical coordinates as usually:
\bea
&&k_3=k \cos\theta
\nn \\
&&k_1=k \sin\theta \cos \phi
\nn \\
&&k_2=k \sin\theta \sin \phi \, .
\eea

Making the aforementioned 2d Wick rotation, taking the Dirac trace on $\Sigma$ and integrating over the angles we get for the function $B$ the same equation we derived directly in Minkowski space, i.e. (\ref{ET1}). 
However now, the equation for $A$ can be more easily derived:
\bea \label{ET2}
A(p)&=&1+i\frac{e^2}{4\pi^2}\int_0^{\infty} d k
\frac{k^2}{p^2} S_v(k^2) \left[-I+(1-\xi)I\right]
\nn \\
I&=&1+\frac{p^2+k^2}{2pl}\ln\left|\frac{k-p}{k+p}\right|
\eea
with the propagator function $S_v$ defined by (\ref{sv}).
We can see that like in the standard ES formulation  we get $A=1$ exactly  in quenched rainbow approximation in Landau gauge $\xi=0$.

\section{Numerical solutions for QED2+1}

In this talk we present Minkowski solution of Eq. (\ref{ET1}) for the  mass function $M=B$.  We do not perform the  continuation to the spacelike axis, which remains to be done in the future. We will consider the nonzero Lagrangian mass  $m$ and interaction strength characterized by  charge $e$.
QED2+1 is  a superrenormalizable theory and  as we have neglected the photon polarization
it turns to be completely ultraviolet finite and  no renormalization is required at all.
 We assume that the imaginary part of the mass function is dynamically generated and to get the  numerical solution  we split the SDE (\ref{ET1}) to the coupled equations for the real and imaginary part of $B$ and solve these  two coupled 
integral equations simultaneously  by the method of  iterations.

The confinement of QED2+1 electron certainly means that we have no free electrons in asymptotic states of any considered process. Furthermore, in quantum theory, physical degrees of freedom are necessarily subject to a probabilistic interpretation implying unitarity and positivity; the physical part of the state space of QCD should be equipped with a positive
(semi-)definite metric. Therefore one way to investigate whether a certain degree of freedom
is confined, is to search for positivity violations in the spectral representation of the
corresponding propagator. 
The standard way in Schwinger-Dyson QCD and QED2+1 studies is to construct  the Schwinger function and check the violation of reflection positivity indirectly from the Euclidean solution. Such  implications of confinement to singularity structure  have been recently studied in the many papers devoted to some approximation of  QCD and particularly  in QED2+1 \cite{BARACORO2008,BRSR2009}. The sufficient condition for the confinement is the absence of Khallen-Lehmann representation for propagator. The obtained solution is complex everywhere with nonzero Im part at the beginning, it is quite obvious that such Minkowski space solution does not satisfy Khallen-Lehmann representation.

To characterize complex mass dynamical generation it is convenient to introduce dimensionless parameter
\be
\kappa=\frac{e^2}{2m}.
\ee
We set up the scale by taking $m=1$ at any units. Pure dynamical chiral symmetry breaking $(m=0)$ is naturally achieved  at large  $\kappa$ limit.

The phase $\phi_M$ of the complex mass function defined by $M=|M|e^{i\phi_M}$
is shown in Fig.1. for a various value of $\kappa$.  For very large $\kappa$ we get the dynamical chiral symmetry breaking in which case the obtained infrared phase  is $\phi_M(p^2=0,\kappa=\infty)=87.5^o$ while more interestingly it vanishes for very small $\kappa$. There is no imaginary part generation for  fermion selfenergy bellow some "critical" value of $\kappa$ , especially one can observe $\phi_M(p^2=0,\kappa<0.0191\pm 0.0001)=0$ for one flavor ladder QED2+1. The absolute value of $M$ is displayed in the right panel of Fig.1. for the various value of the coupling $ \kappa$. As $\kappa$ decreases the expected complex singularities gradually moves from complex plane to the real axis and the function develops  something like known threshold enhancement. So for heavy electron we can have a real pole, albeit confinement.

\begin{figure}
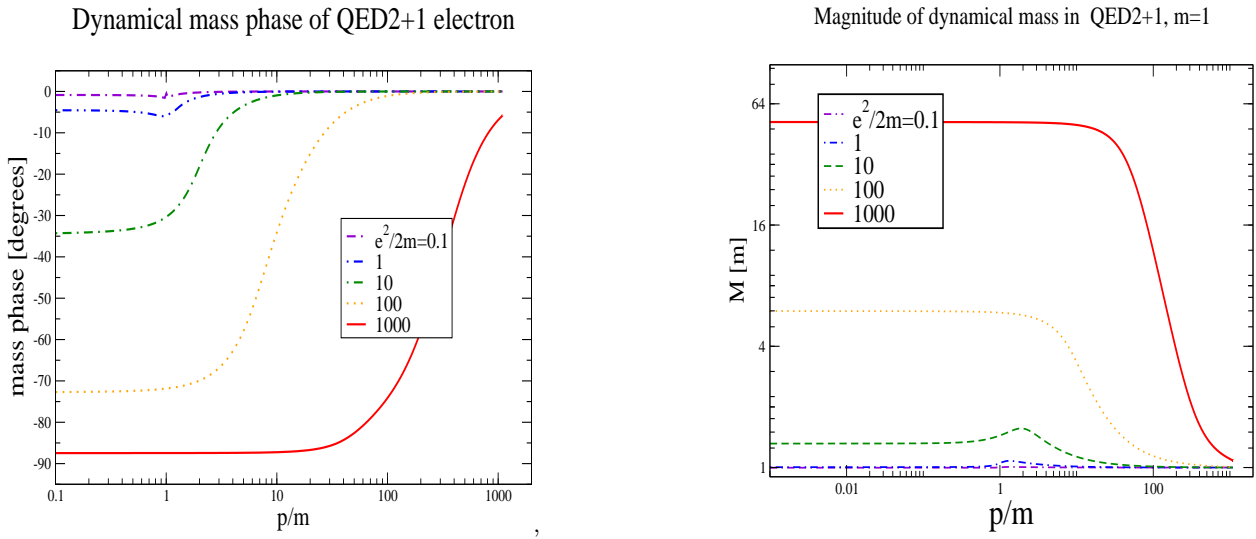

\centerline{\hspace{-1.0cm}\epsfig{figure=uhlyqed3.eps,width=7truecm,height=7truecm,angle=0},\hspace{+2.5cm}\epsfig{figure=magnqed3.eps,width=7truecm,height=7truecm,angle=0}}
\caption[caption]{Left:Phase $\phi $ of the dynamical mass function $M=|M|e^{i\phi}$ of electron living in 2+1 dimensions for different $\kappa$, scale is $m=1$.Right:Magnitude $|M|$ of the running mass $M=|M|e^{i\phi}$ of electron living in 2+1 dimensions for different $\kappa$, scale is $m=1$.} 
\end{figure}

\section{Numerical solutions for quark SDE }

Exactly in the same manner the quark gap equation has been solved in two models partially differed by the modeled 
gluon propagator. In the ladder approximation the quark SDE reads
\bea \label{gap2}
S^{-1}(p)&=&S^{-1}_0(p)-\Sigma(p) \, ,
\nn \\
\Sigma(p)&=&i C_A g^2\int\frac{d^4q}{(2\pi)^4}\gamma_{\alpha}
G^{\alpha\beta}(p-q)S(q)\gamma_{\beta} \, ,
\eea
When compared to QED2+1, the main difference is the number of spatial dimension here. Transforming the equation 
(\ref{gap2}) to Temporal Euclidean space, then the integral  kernels are purely real if no dynamical generation of imaginary mass would appear.  The calculation is performed in the Landau gauge, where the gluon propagator is transverse. The model I of the paper \cite{Ja} assumes perturbative pole in gluon  propagator $G^{\alpha\beta}$ modified by the effective running coupling, while the full gluon form factor has been modeled by spectral representation in model II, so the propagator is with branch point instead of the pole at zero momenta. Some other details of the models can be find in the paper \cite{Ja}. The infrared  
behaviour of the functions $M$ is shown in Fig2. The left panel shows the magnitude $|M|$ of the running quark mass function $M=|M|e^{i\phi}$ for modeled QCD I,II and their chiral limit CI,CII. The Right panel shows the phase $\phi$ of the running quark mass function $M=|M|e^{i\phi}$ for the models I and II respectively, axis momentum is in the units of  $\Lambda_{QCD}$.

\begin{figure}
\centerline{\hspace{-1.0cm}\epsfig{figure=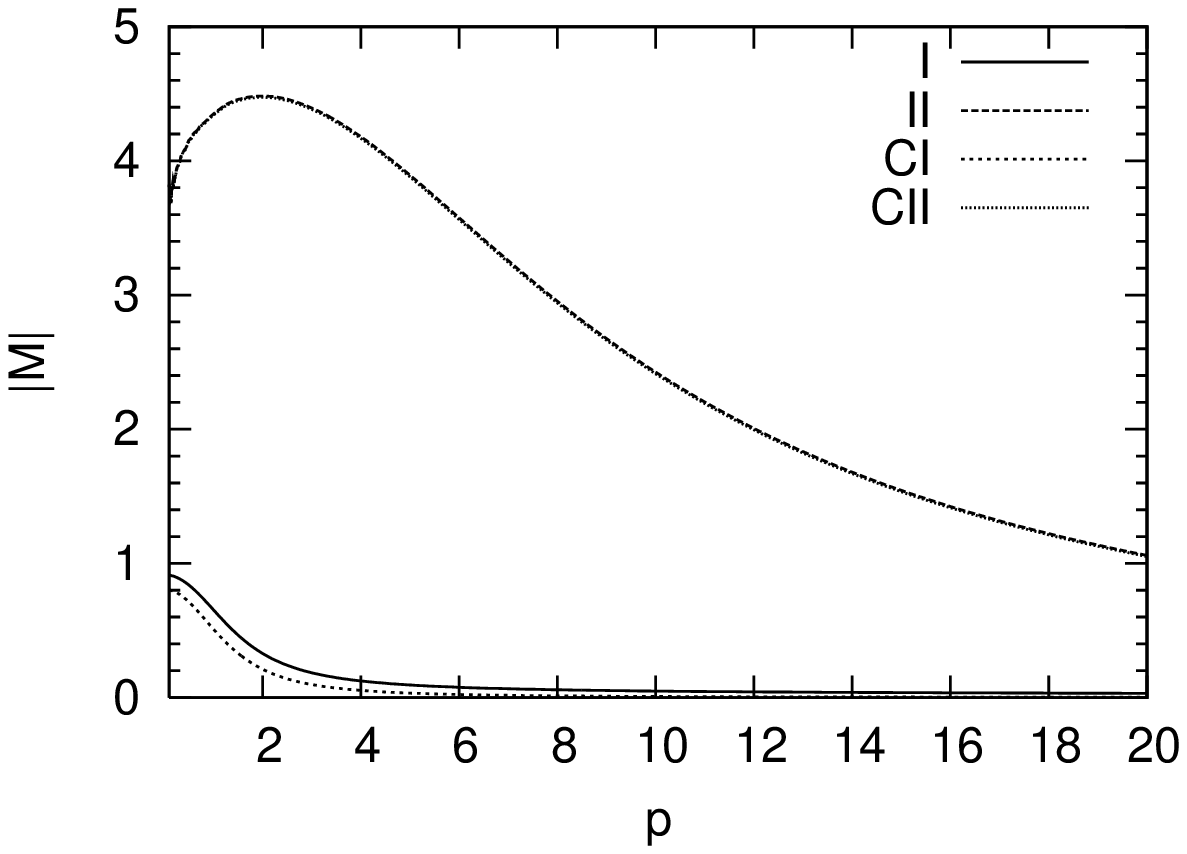,width=7truecm,height=7truecm,angle=0},\hspace{+2.5cm}\epsfig{figure=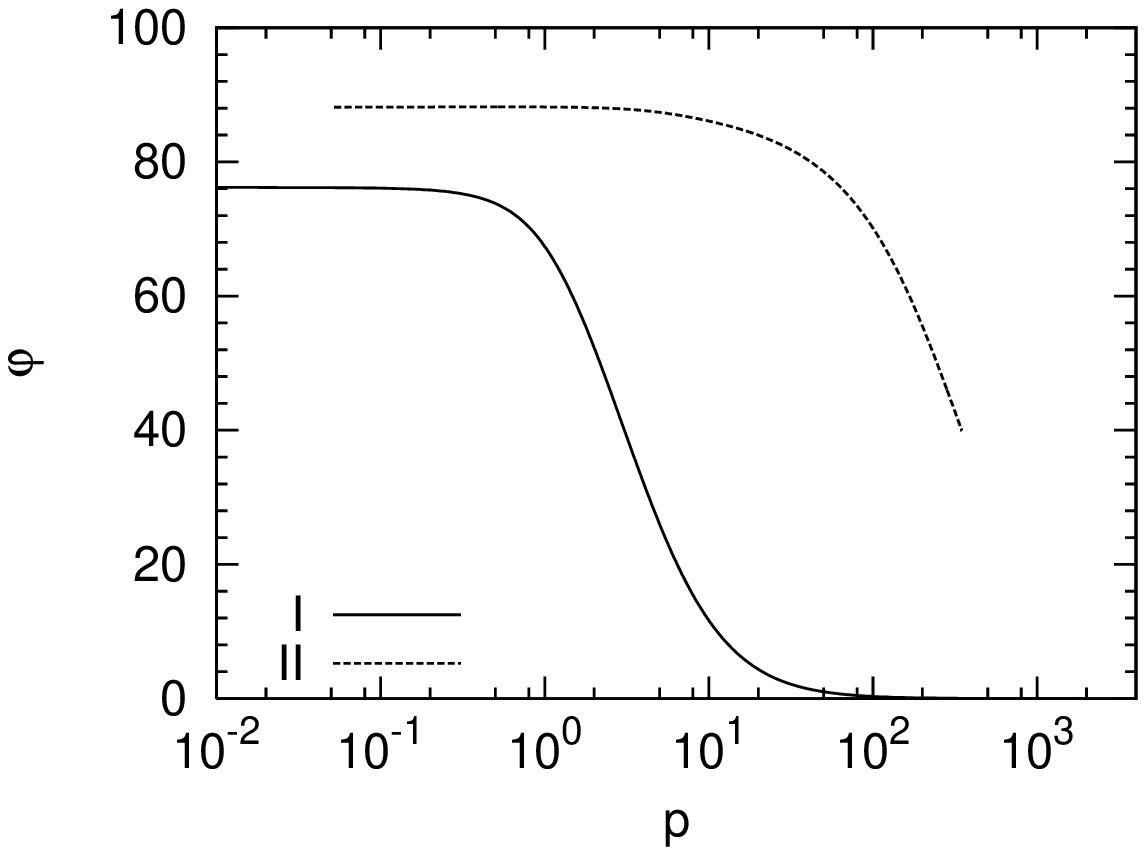,width=7truecm,height=7truecm,angle=0}}
\caption[caption]{Running quark masses as described in the text} 
\end{figure}

\section{Summary and conclusions}

We have presented the first analysis of the  electron gap equation  
in  Minkowski  space. 
The other result, although based on the simple
ladder approximation in given gauge, is the proof of the exact equivalence between  the theories defined in Minkowski 2+1 and 3D Temporal Euclidean space. No similar is known about  the standard Euclidean formulation and its relation with spacelike  subspace of  Minkowski space. We clearly argue  -the well known calculational trick in quantum theory- is based on an unjustified assumption in the case of QED2+1. 

Consequently the Euclidean framework has been applied to 
QCD light quarks.
In both models,QED2+1 and QCD, the explicit or the  dynamical generation of  imaginary part of the fermion mass  leads to the absence of Khallen-Lehmann representation, providing thus confining solution. Minkowski QED2+1 has been shown to exhibit spontaneous chiral symmetry breaking -the mass function  has nontrivial solution in the limit $m=0$. The Temporal Euclidean space,  ad hoc introduced for 3+1 dimensional QCD  opens up a variety of questions. If there is any, what is the relation between Euclidean spaces (spacelike and timelike ones) formulation and the original Minkowski space? The question of equivalence is much more difficult to answer here  since most of Minkowski integrals are badly defined. It stays for future investigation.


\end{document}